\documentstyle[amsmath,a4,11pt,epsfig]{article}
\def\slash{\!\!\!/}
\begin{document}     
\title{{\bf Power scaling rules for charmonia production and HQEFT}
\vspace*{0.2in}
\thanks{Research partially supported by CICYT under contract AEN99-0692}}
\author{Miguel-Angel Sanchis-Lozano \thanks{E-mail: mas@evalo1.ific.uv.es} \\
\\
\it Departamento de F\'{\i}sica Te\'orica and IFIC\\
\it Centro Mixto Universidad de Valencia - CSIC\\
\it 46100 Burjassot, Valencia, Spain}
\maketitle 
\vspace*{1.0in}
\abstract{We discuss the power scaling rules
along the lines of a complete Heavy Quark Effective Field
Theory (HQEFT) for the description of heavy quarkonium
production through a color-octet mechanism. To this end, 
we firstly derive a tree-level heavy quark effective Lagrangian 
keeping both particle-antiparticle mixed sectors allowing for
heavy quark-antiquark pair annihilation and
creation, but describing only low-energy modes
around the heavy quark mass. Then we show the consistency of using HQEFT
fields in constructing four-fermion local operators {\em \`a la NRQCD},
to be identified with standard color-octet matrix elements. We analyze some
numerical values extracted from charmonia production
by different authors and 
their hierarchy in the light of HQEFT.}
\large
\vspace{-17.cm}
\begin{flushright}
  FTUV-01-0313 \\
  IFIC/01-09 \\
  hep-ph/0103140
\end{flushright} 
\vspace{17.cm}
\newpage
\section{Introduction}

The application of effective theories directly derived from
first principles - 
Quantum Chromodynamics (QCD) in particular - to heavy flavor
physics has currently become a very important
tool to cope with the complexity of the strong interaction
dynamics. One of its advantages is the
appearance of symmetries in the very large quark
mass limit, not manifest in
the full theory. Moreover, symmetry breaking effects due to
the finiteness of quark masses can be systematically 
incorporated via power expansions in the reciprocal of the
heavy quark mass. One of the foremost applications of
heavy-quark spin-flavor symmetry
has led to accurate and almost model-independent
extractions of some Kobayashi-Maskawa matrix elements 
in inclusive and exclusive $B$ meson decays (see \cite{neub2} for a review).

Over the nineties, a heavy quark effective
theory (HQET) was developed 
\cite{grins,neub} to deal with the phenomenology of
heavy-light mesons and baryons. 
Moreover, there were early attempts
to extend its initial scope to heavy-heavy 
bound states like the $J/\psi$ resonance \cite{mas93} or the $B_c$ meson 
and doubly heavy baryons \cite{mas95}. 
On the other hand, a close
approach also emerged in parallel, especially designed to describe
heavy quarkonia inclusive decay and production, the 
Non-Relativistic QCD (NRQCD)\cite{bodwin,lepage}.  In this
formalism, the decay widths or cross sections are factorized
as a sum of products of short-distance coefficients
and long-distance NRQCD matrix elements.

Although both HQET and NRQCD Lagrangians present
a formal analogy,  
there is a basic difference between both frameworks. 
In the former one, there are basically two relevant scales: the heavy
quark mass $m_Q$ and ${\Lambda}_{QCD}$, the
characteristic parameter of the strong interaction. In NRQCD, besides 
${\Lambda}_{QCD}$, there are (at least) another two   
low-energy scales: $m_Qv'$ (soft) and $m_Q{v'}^2$ (ultrasoft) 
\footnote{In this paper we shall employ $v'$ to denote
the three-velocity of heavy quarks inside quarkonium while
$v$ will stand for the hadron four-velocity, as usual
in HQET; since residual momenta of heavy quarks will not be
neglected in the hadron rest frame, $v'$ obviously
does not exactly coincide with the spatial
component of $v$, i.e. $v'\ {\neq}\ {\mid}{\bf v}{\mid}$.}.

Recently, several applications of HQET involving both heavy quark
and antiquark fields altogether to the study of 
the phenomenology of the weak and strong interactions
have appeared in the literature
(\cite{wu,wu1} and \cite{mas00}, respectively).
At the same time it becomes suitable
a careful survey of the foundations on which such a 
theoretical framework is based. Therefore, in this paper
firstly we shall focus on the quantum aspects of
the heavy quark effective fields,
summarizing some work
presented in previous references \cite{mas97,mas97b}.
In particular, we shall keep both particle-antiparticle
mixed sectors of the effective Lagrangian, in
accordance with the
pioneering work by Wu in Ref. \cite{wu93},  
further developed and discussed in \cite{mas99}. Hereafter we 
shall refer to this
extended framework as HQEFT as in \cite{wu}, to distinguish
it from $\lq\lq$usual'' HQET.
 
Furthermore, one of the goals of this paper is to analyze whether
HQEFT could provide an appropriate framework to describe
meaningfully the
hadroproduction of charmonium.  In particular, we shall analyze 
the power counting of four-fermion local operators, whose scaling rules
for charmonia have been recently hypothesized
in Ref. \cite{leibovich} along the lines of HQET.
We shall examine this issue with the help of diverse
extractions of some matrix elements performed by several authors.

\section{Notation and definitions}

Let us start following Ref. \cite{mas97} by writing the 
plane wave Fourier expansion for a
fermionic field $Q_v(x)$ corresponding to an almost on-shell
heavy quark or antiquark (we omit indices relative to flavor 
and color) moving inside
a hadron with four-velocity $v$, 
\begin{equation}
Q_v(x) = Q_v^{(+)}(x)+
Q_v^{(-)}(x) = \int \frac{d^3{\bf p}}{J}\ \sum_{r}\ 
[\ b_r({\bf p})\ u_r({\bf p})\ e^{-ip{\cdot}x}+ 
\ \tilde{b}_r^{\dag}({\bf p})\ 
v_r({\bf p})\ e^{ip{\cdot}x}\ ]
\end{equation}
where $r$ refers to the spin and $J$ stands for the chosen
normalization; $b_r({\bf p})/\tilde{b}_r^{\dag}({\bf p})$
is the annihilation/creation operator for a heavy quark/antiquark
with three-momentum ${\bf p}$ ($p^0\ {\simeq}+\sqrt{m^2+{\bf p}^2}$). 
Let us firstly focus on the particle sector of the \vspace{0.2in} theory.
\newline
{\bf Particle \vspace{0.1in} Sector}
\par
As the heavy quark is almost on-shell one should require for
{\em each Fourier component} that
$p^2=m_Q^2+{\Delta}^2$
where ${\Delta}^2$ satisfies
${\lim}_{m_Q{\rightarrow}\infty}{\Delta}^2/m_Q^2\ {\rightarrow}\ 0$. 
From now on, we shall set ${\Delta}^2=0$ as a good approximation
for heavy quarks bound in hadrons.

On the other hand, spinors are normalized such that
$u_r^{\dag}({\bf p})u_s({\bf p})=2p^0N\ {\delta}_{rs}$ and the 
creation/annihilation operators satisfy:
\begin{equation} 
[b_r({\bf p'}),b_s^{\dag}({\bf p})]_+=
K\ {\delta}_{rs}\ {\delta}^3({\bf p'}-{\bf p})
\end{equation}
where $K,\ N$ are the corresponding normalization factors.

Now let us redefine the momentum of each Fourier component in Eq. (1)
according to HQET as the sum of a mechanical part and
a {\em Fourier residual four-momentum} $k$, 
\begin{equation}
p\ =\ m_Qv\ +\ k
\end{equation}
\par
Hence one can write from (1)  
\begin{equation}
Q_v^{(+)}(x)\ =\ e^{-im_Qv{\cdot}x}\ \int\  
\frac{d^3{\bf k}}{J}\ \sum_r\ b_r(v,{\bf k})\ u_r({\bf k})\ 
e^{-ik{\cdot}x}
\end{equation}
\par
The main point to be stressed again is that we shall require that each
Fourier component should satisfy the almost on-shell condition
$p_Q^2\ {\simeq}\ m_Q^2$. 
Therefore, 
\begin{equation}
k^2\ +\ 2m_Qv{\cdot}k\ =\ 0
\end{equation}

Thus, in the hadron rest frame, the kinetic energy
of the heavy quark $v{\cdot}k=k^0$ is related to
${\bf k}$ through the constraint 
\begin{equation}
(k^0)^2+2m_Qk^0-{\bf k}^2\ =\ 0
\end{equation}
which yields the expected relation for the positive root
$k^0=\sqrt{m_Q^2+{\bf k}^2}-m_Q$.
In the non-relativistic limit obviously, 
$k^0\ {\simeq}\ {\bf k}^2/2m_Q$.

Notice that the annihilation (and creation) operators and spinors 
have been simply {\em relabeled} in Eq. (4), satisfying the same 
normalization as above, though expressed in terms of the Fourier 
residual momentum ${\bf k}$. 
In particular, $b_r(v,{\bf k})/b_r^{\dag}(v,{\bf k})$
correspond to annihilation/creation operators
of a heavy quark with residual momentum ${\bf k}$ in a hadron
moving with four-velocity  $v$, satisfying accordingly
\begin{equation}
[b_r(v,{\bf k}'),b_s^{\dag}(v,{\bf k})]_+=
K\ {\delta}_{rs}\ {\delta}^3({\bf k}'-{\bf k})
\end{equation}
and
\begin{equation}
{\sum}_r\ u_r({\bf k})\bar{u}_r({\bf k})\ =\  N\ 
[\ m_Q(1+v{\slash})+k{\slash}\ ] 
\end{equation}
where the normalization factors must obey the combined relation:  
\cite{dona}
\footnote{
Note that $I{\neq}1/(2\pi)^32k^0$ since $k$ and $p$ are
not related through a Lorentz transformation; indeed, Eq. (3)
represents an energy-momentum shift, to get rid of the unwanted
frequency modes to be subsequently
eliminated in the effective theory.}

\begin{equation} 
I\ =\ \frac{K\ N}{J^2}\ =\ \frac{1}{(2\pi)^3\ 2p^0}\ =\ 
\frac{1}{(2\pi)^3\ 2(m_Qv^0+k^0)}
\end{equation}

On the other hand, it is quite usual in the literature to 
identify effective heavy quark fields with those ${\lq}{\lq}$leading"
components of the Fourier expansion corresponding to momenta
$p^{\mu}\ {\simeq}\ m_Qv^{\mu}$ (or equivalently with $k^{\mu}$ components
close to zero), so they rather look like single spinors or anti-spinors
at leading order in $1/m_Q$. 
However, this is a too restrictive view since
in constructing HQEFT one should allow
the ${\bf k}$ components of the effective quantum fields to range 
over values of the order of ${\Lambda}_{QCD}$.

Next let us introduce the effective
fields in the following standard manner \cite{neub} \cite{mannel},

\begin{equation}
h_v^{(+)}(x)\ =\ e^{im_Qv{\cdot}x}\ \frac{1+v{\slash}}{2}\ 
Q_v^{(+)}(x)
\end{equation}
\begin{equation}
H_v^{(+)}(x)\ =\ e^{im_Qv{\cdot}x}\ \frac{1-v{\slash}}{2}\ 
Q_v^{(+)}(x)
\end{equation}
where $H_v^{(+)}(x)$ represents
the ${\lq}{\lq}$small" (lower in the hadron reference frame) 
component of the corresponding spinor field. Thus one may identify
from the expansion (4)
\begin{equation}
h_v^{(+)}(x)\ =\  \frac{1+v{\slash}}{2}\ \int\ \frac{d^3{\bf k}}{J}
\ \sum_{r}\ b_r(v,{\bf k})\ u_r({\bf k})\ 
e^{-ik{\cdot}x}
\end{equation}
\begin{equation} 
H_v^{(+)}(x)\ =\ \frac{1-v{\slash}}{2}\ \int\
\frac{d^3{\bf k}}{J}\ \sum_{r}\ b_r(v,{\bf k})\ 
u_r({\bf k})\ e^{-ik{\cdot}x} 
\end{equation}  

Sometimes it can be read in
the literature that $h_v^{(+)}$ annihilates a heavy quark 
whereas allegedly $H_v^{(+)}$ creates a heavy antiquark, both with velocity 
$v$. Although $H_v^{(+)}$ may take into account the antiparticle
features of a relativistic fermion, its Fourier decomposition 
does not allow for such a process, as can be seen from (13).

\section{Anti-particle Sector}

Proceeding in a parallel way as in the particle sector, let us remark
however that now the Fourier expansion of $Q_v^{(-)}(x)$
involves negative frequencies. Therefore, starting from
Eq. (1) we introduce the Fourier residual momentum in this case as
\begin{equation}
p\ =\ m_Qv\ -\ k
\end{equation}
so $k^0$ will explicitly exhibit its negative sign. 
In effect, let us write
\begin{equation}
Q_v^{(-)}(x)\ =\ 
\ e^{im_Qv{\cdot}x}\ \int\   
\frac{d^3{\bf k}}{J}\ \sum_r\ \tilde{b}_r^{\dag}({\bf k})\ 
v_r({\bf k})\ e^{-ik{\cdot}x}
\end{equation}
\par
The slightly off-shellness condition $p^2\ {\simeq}\ m_Q^2$
now implies from Eq. (14):
\begin{equation}
k^2\ -\ 2m_Qv{\cdot}k\ =\ 0
\end{equation}
which can be written in the frame moving with velocity $v$ as
\begin{equation}
(k^0)^2\ -\ 2m_Qk^0\ -\ {\bf k}^2\ =\ 0
\end{equation}
whose negative root reads
$k^0\ =\ m_Q\ -\ \sqrt{m_Q^2+{\bf k}^2}\ {\simeq}\ -{\bf k}^2/2m_Q$.

The effective fields are then defined as
\begin{equation}
h_v^{(-)}(x)\ =\ e^{-im_Qv{\cdot}x}\ \frac{1-v{\slash}}{2}\ 
Q_v^{(-)}(x)=\  \frac{1-v{\slash}}{2}\ \int\ \frac{d^3{\bf k}}{J}
\ \sum_{r}\ \tilde{b}_r^{\dag}({\bf k})\ v_r({\bf k})\ 
e^{-ik{\cdot}x}       
\end{equation}
\begin{equation}
H_v^{(-)}(x)\ =\ e^{-im_Qv{\cdot}x}\ \frac{1+v{\slash}}{2}\ 
Q_v^{(-)}(x)\ =\ \frac{1+v{\slash}}{2}\ \int\
\frac{d^3{\bf k}}{J}\ \sum_{r}\ \tilde{b}_r^{\dag}({\bf k})\ 
v_r({\bf k})\ e^{-ik{\cdot}x} 
\end{equation}
\vspace{0.1in}
where the anti-spinors satisfy
\begin{equation}
{\sum}_r\ v_r({\bf k})\bar{v}_r({\bf k})\ =\  
 N\ [\ m_Q(v{\slash}-1)-k{\slash}\ ] 
\end{equation}

\section{A complete tree-level HQEFT Lagrangian}

Once introduced the notation and 
definitions for the heavy-quark effective fields in the
previous Section, we actually get started by
expressing the Lagrangian in terms of the effective fields 
keeping all non-null terms, leading in principle 
to the possibility of describing
annihilation or creation of heavy quark-antiquark
pairs at tree level. Indeed after performing an
energy-momentum shift by introducing a center-of-mass residual
momentum, only low energy modes of the fields (about the heavy quark mass) 
would be involved making meaningful our approach within the framework of 
an effective theory.
 
Thereby a crucial difference of this approach w.r.t. other standard works is 
that we are concerned with the existence of two-fermion terms in the
transformed  Lagrangian mixing large components of the 
heavy quark and heavy antiquark fields, 
i.e. $h_v^{({\pm})}{\Gamma}h_v^{({\mp})}$, 
where ${\Gamma}$ stands for a combination of Dirac gamma matrices
and covariant derivatives. The implications on the description
of heavy quarkonia production will be seen in Section 6.

\def\slash{\!\!\!\!/}

The tree-level QCD Lagrangian is our point of departure: 
\begin{equation}
{\cal L}\ =\ \overline{Q}\ (i\overrightarrow{D{\slash}}-m_Q)\ Q
\end{equation}
where
\begin{equation}
Q\ =\ Q^{(+)}+Q^{(-)}\ =\ 
e^{-im_Qv{\cdot}x}\biggl[\ h_v^{(+)}+H_v^{(+)}\ \biggr]
+e^{im_Qv{\cdot}x}\biggl[\ h_v^{(-)}+H_v^{(-)}\ \biggr]
\end{equation}
and $D$ standing for the covariant derivative
\[
\overrightarrow{D}^{\mu}\ =\ \overrightarrow{\partial}^{\mu}\ -\ igT_aA_a^{\mu}
\]
with $T_a$ the generators of $SU(3)_c$. Substituting (22) into (21)
one easily arrives at
\begin{equation}
{\cal L}\ =\ {\cal L}^{(++)}\ +\ {\cal L}^{(--)}\ +\ {\cal L}^{(-+)}\ +\ 
{\cal L}^{(+-)}
\end{equation}
where we have explicitly splitted the Lagrangian into four different
pieces corresponding to the particle-particle, antiparticle-antiparticle
and both particle-antiparticle sectors. The former one has the form   
\begin{equation}
{\cal L}^{(++)}=\bar{h}_v^{(+)}iv{\cdot}\overrightarrow{D}h_v^{(+)}-
\bar{H}_v^{(+)}(iv{\cdot}\overrightarrow{D}+2m_Q)H_v^{(+)}
+\bar{h}_v^{(+)}i\overrightarrow{D{\slash}}_{\bot}H_v^{(+)}+\bar{H}_v^{(+)}i\overrightarrow{D{\slash}}_{\bot}h_v^{(+)}
\end{equation}
corresponding to the usual HQET Lagrangian still containing
both $h_v^{(+)}$ and $H_v^{(+)}$  fields. We employ the common notation where
perpendicular indices are implied according to
\[ D^{\mu}_{\bot}\ =\ D_{\alpha}\ (g^{\mu\alpha}-v^{\mu}v^{\alpha}) \]
\par
Regarding the antiquark sector of the theory
\begin{equation}  
{\cal L}^{(--)}=-\bar{h}_v^{(-)}iv{\cdot}\overrightarrow{D}
h_v^{(-)}+\bar{H}_v^{(-)}(iv{\cdot}\overrightarrow{D}-2m_Q)H_v^{(-)}
+\bar{h}_v^{(-)}i\overrightarrow{D{\slash}}_{\bot}H_v^{(-)}+
\bar{H}_v^{(-)}i\overrightarrow{D{\slash}}_{\bot}h_v^{(-)}
\end{equation}
\par
The latter expressions, considered as quantum field Lagrangians, 
certainly do 
not afford tree-level heavy quark-antiquark pair
creation or annihilation processes stemming from the terms
mixing ${h}_v^{(\pm)}$ and ${H}_v^{(\pm)}$ fields since they contain
either annihilation and creation operators of heavy quarks
or annihilation and creation operators of heavy antiquarks
separately.

Nevertheless there are two extra pieces in the Lagrangian (23):
\begin{eqnarray}
& {\cal L}^{(-+)} & =  e^{-i2m_Qv{\cdot}x}\ {\times} \\
& &
[\ \bar{H}_v^{(-)}iv{\cdot}\overrightarrow{D}h_v^{(+)}
-\bar{h}_v^{(-)}(iv{\cdot}\overrightarrow{D}+2m_Q)H_v^{(+)}
+\bar{h}_v^{(-)}i\overrightarrow{D{\slash}}_{\bot}h_v^{(+)}+
\bar{H}_v^{(-)}i\overrightarrow{D{\slash}}_{\bot}H_v^{(+)}\ ] \nonumber
\end{eqnarray}
and
\begin{eqnarray}
& {\cal L}^{(+-)} & =  e^{i2m_Qv{\cdot}x}\ {\times} \\
& & 
[\ -\bar{H}_v^{(+)}iv{\cdot}\overrightarrow{D}h_v^{(-)}
+\bar{h}_v^{(+)}(iv{\cdot}\overrightarrow{D}-2m_Q)H_v^{(-)}
+\bar{h}_v^{(+)}i\overrightarrow{D{\slash}}_{\bot}h_v^{(-)}+
\bar{H}_v^{(+)}i\overrightarrow{D{\slash}}_{\bot}H_v^{(-)}\ ] \nonumber
\end{eqnarray}
where use was made of the orthogonality of the $h_v^{(\pm)}$ and 
$H_v^{(\pm)}$ fields \cite{mas99}.
As could be expected, there are indeed pieces mixing both 
quark and antiquark fields
leading to the possibility of annihilation/creation processes.
After all, Lagrangian (23) is
still equivalent to full (tree-level) QCD 
\footnote{Superscript ${\lq\lq}(+)/(-)$" on the effective fields labels
the particle/antiparticle (i.e. heavy quark/antiquark)
sector of the theory \cite{georgi}. Actually 
$\bar{h}_v^{(+)}\ (\bar{h}_v^{(-)})$ corresponds to
negative (positive) frequencies associated with 
creation (annihilation) operators of quarks (antiquarks).
In fact some extra ${\lq\lq}+/-$" signs should be added on the 
conjugate fields, 
i.e. $\bar{h}_v^{(+)-}$ and $\bar{H}_v^{(-)+}$, which
however will be omitted to shorten the notation.}. 

Let us remark that, at first sight, one might think that the rapidly 
oscillating exponentials in Eqs. (26-27) would make both ${\cal L}^{(-+)}$ and
${\cal L}^{(+-)}$ pieces to vanish, once integrated over all velocities
according to the most general Lagrangian \cite{georgi}.
However, notice that actually this should not be the case for momenta of 
order $2m_Qv$ of the gluonic field present in the covariant derivative. 
In fact, only such high-energy modes would survive, corresponding to the 
physical situation on which we are focusing, i.e. heavy quark-antiquark 
pair annihilation or creation.

From Eq. (26), the heavy quark-gluon coupling for an annihilation 
process is given by 
\begin{eqnarray}
{\cal L}_{coupling}^{(-+)} &  & =e^{-i2m_Qv{\cdot}x}\ gT_aA_{\mu}^a\ {\times} \nonumber \\
& [ & \bar{H}_v^{(-)}\ v^{\mu}\ h_v^{(+)}\ \ \ \ \ \ \ \ (a) \nonumber \\
& - & \bar{h}_v^{(-)}\ v^{\mu}\ H_v^{(+)}\ \ \ \ \ \ \ \ (b) \nonumber \\
& + & \bar{h}_v^{(-)}\ \gamma_{\bot}^{\mu}\ h_v^{(+)}\ \ \ \ \ \ \ \ \ 
(c) \nonumber \\
& + & \bar{H}_v^{(-)}\ \gamma_{\bot}^{\mu}\ H_v^{(+)}\ ]\ \ \ \ \ (d)  
\end{eqnarray}

There is an equivalent expression coming from 
${\cal L}_{coupling}^{(+-)}$  just by means of the
substitution $v{\rightarrow}-v$, corresponding to
a heavy quark-antiquark pair creation process.

Let us make an important remark concerning the gluonic field.
As can be seen at once, the exponential factor $e^{-i2m_Qv{\cdot}x}$ in
Eq. (28) cancels against the strong $x$ dependence of 
the $A_{\mu}^a$ field
creating a hard gluon. Therefore we can write
\begin{equation}
A_{\mu}^{soft}(x)=e^{-i2m_Qv{\cdot}x}\ A_{\mu}(x)
\end{equation}

Therefore, once removed the strong dependence of the gluon momentum on
(twice) the heavy quark mass, we are left with an effective gluonic field
$A_{\mu}^{soft}$, with a mild dependence on the residual momentum,
similarly as for heavy quark effective fields.
Thus, the physical description of the annihilation process
shown in Fig. 1 could be performed in terms of massless quarks
and a soft gluonic field. This is pretty similar to the
situation reached in heavy-light systems, constituting
a cornerstone in our approach
to be developed in the following Sections, providing a theoretical
basis for the use of HQEFT to describe heavy quarkonium production.

\begin{figure}[htb]
\centerline{
\epsfig{figure=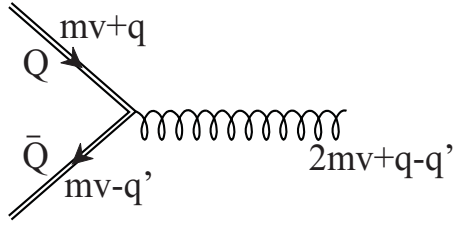,height=3.cm,width=6.cm}}
\caption{\small{Heavy quark-antiquark pair annihilation into
a gluon of momentum $2m_Qv+q-q'$. In the center-of mass system
$q=(q_0,\bf{q})$ and $q'=(-q_0,{\bf q})$.
This process can be described meaningfully by HQEFT if
${\bf q^2}<<m_Q^2$, i.e. the square invariant mass of the gluon
is close to $4m_Q^2$.}}
\end{figure}

Next we want to eliminate the unwanted degrees of freedom associated to
the $\lq\lq$small'' components $H_v^{(\pm)}$ in $(a)$, $(b)$ and $(d)$
in Eq. (28). 
Notice that the piece labeled as (c) is the leading one 
in the above development.

\def\slash{\!\!\!/}

\section{Derivation of the annihilation vertex}

In all our later development we shall assume almost free
quarks in the initial (or final) state (see Fig. 1). Hence, heavy
quark fields appearing on the external legs of the Feynman diagram
should be solutions of the unperturbed Dirac equation of motion.
Hence we can write
\begin{eqnarray}
H_v^{(+)} & = & (iv{\cdot}\overrightarrow{\partial}+2m_Q-i\epsilon)^{-1}
i\overrightarrow{\partial{\slash}}_{\bot}h_v^{(+)} \\
H_v^{(-)} & = & (-iv{\cdot}\overrightarrow{\partial}+2m_Q-i\epsilon)^{-1}
i\overrightarrow{\partial{\slash}}_{\bot}h_v^{(-)}
\end{eqnarray}
and for the conjugate fields
\begin{eqnarray}
\bar{H}_v^{(+)} & = & 
\bar{h}_v^{(+)}i\overleftarrow{\partial{\slash}}_{\bot}
(iv{\cdot}\overleftarrow{\partial}-2m_Q+i\epsilon)^{-1} \\
\bar{H}_v^{(-)} & = & 
\bar{h}_v^{(-)}i\overleftarrow{\partial{\slash}}_{\bot}
(-iv{\cdot}\overleftarrow{\partial}-2m_Q+i\epsilon)^{-1}
\end{eqnarray}

Let us note that we are using the free particle equations 
which can be viewed as derived from the non-interacting parts of
the Lagrangians ${\cal L}^{(++)}$ and ${\cal L}^{(--)}$
respectively, i.e. we
neglect the soft gluon interaction among heavy quarks amounting to 
describe them as plane waves, i.e. actually no bound states
as a first approximation \cite{hussain}.

In momentum space we will write 
\begin{equation}
u(m_Qv+q)=\biggl[\ 1+\frac{q{\slash}_{\bot}}{2m_Q+v{\cdot}q}\ \biggr]\ u_h(q)
\end{equation}
\begin{equation}
\overline{v}(m_Qv-q')=\overline{v}_h(-q')\ 
\biggl[\ 1-\frac{q{\slash}'_{\bot}}{2m_Q-v{\cdot}q'}\ \biggr]
\end{equation}
where $u(p)$ denotes the full QCD spinor  
whereas $u_h(q)$ represents the HQET spinor i.e. obeying 
$v{\slash}u_h=u_h$; similarly $v(p)$ denotes the full QCD 
antispinor and $v_h(-q')$) represents the HQET antispinor obeying 
$v{\slash}v_h=-v_h$.

Actually, in order to arrive to Eqs. (34-35) from Eqs. (30-33)
denominators have to be expanded as power series of derivatives
acting on the $x$-dependent factors, assumed exponentials,
to be finally resummed as geometric series of ratio
 $v{\cdot}q/2m_Q=-v{\cdot}q'/2m_Q=-q^2/4m_Q^2$ according to the on-shell 
conditions (5) and (16).
As a consequence, Eqs. (34-35) are only valid under the condition 
$-q^2<4m_Q^2$, which implies ${\bf{q}^2}<8m_Q^2$. Therefore, the
requirement ${\bf{q}^2}<m_Q^2$ satisfies the above condition, 
allowing a non-relativistic expansion.

From the ${\cal L}^{(-+)}$
Lagrangian we readily get for the on-shell heavy quark (vector) 
current coupling to a gluon, suppressing color indices and matrices,
\begin{equation}
\overline{v}_h\ \biggl[{\gamma}^{\mu}_{\bot}+
\frac{q{\slash}'_{\bot}-q{\slash}_{\bot}}{2m_Q+v{\cdot}q}v^{\mu}
+\frac{q{\slash}'_{\bot}{\gamma}^{\mu}q{\slash}_{\bot}}
{(2m_Q+v{\cdot}q)^2}\biggr]\ u_h 
\end{equation}
which also can be written as

\begin{equation}
\overline{v}_h\ \biggl[{\gamma}^{\mu}_{\bot}+
\frac{i{\sigma}^{\mu\nu}(q'_{\bot}-q_{\bot})_{\nu}}{2m_Q+v{\cdot}q}
+\frac{q{\slash}'_{\bot}{\gamma}^{\mu}q{\slash}_{\bot}}
{(2m_Q+v{\cdot}q)^2}\biggr]\ u_h 
\end{equation}
since \cite{mas99}
\[
P_-\ (v^{\mu}{\gamma}^{\nu})\ P_+\ 
=\ P_-\ (i{\sigma}^{\mu\nu}\ +\ 
{\gamma}^{\mu}v^{\nu})\ P_+
\]
with the projectors $P_{\pm}=(1{\pm}v{\slash})/2$, and 
$q'_{\bot\nu}v^{\nu}=q_{\bot\nu}v^{\nu}=0$.

Equations (36-37) are Lorentz covariant and hence valid
in any reference frame. In the following Section we
focus on the center-of-mass system, where some simplifications
occur.

\subsection{Center-of-mass frame}

We shall make use of the anticommutation relation:
\[ q{\slash}'_{\bot}{\gamma}^{\mu}q{\slash}_{\bot}=
2{q'}^{\mu}_{\bot}q{\slash}_{\bot}-
{\gamma}^{\mu}q{\slash}'_{\bot}q{\slash}_{\bot} \] 

Now, in the center-of-mass frame we can write
$q{\slash}'_{\bot}q{\slash}_{\bot}=-{\bf q^2}$,
leading to the expression for the heavy quark vector current
\begin{equation}
\overline{v}_h\ \biggl[\frac{2E_q}{E_q+m_Q}{\gamma}^{\mu}_{\bot}+
\frac{2{q'}^{\mu}_{\bot}q{\slash}_{\bot}}{(E_q+m_Q)^2} \biggr]\ u_h
\end{equation}
where $E_q=m_Q+q^0=\sqrt{m_Q^2+{\bf q}^2}$. Moreover, the 
following identity is 
satisfied (in the Dirac representation):
\begin{equation}
P_-\ q{\slash}_{\bot}\ P_+ = 
\left(
     \begin{array}{ll}
      \ 0 & 0 \\
      {\bf \sigma}{\cdot}{\bf q} & 0     
     \end{array}
\right)
\end{equation}

Therefore identifying $u_h=\sqrt{E_q+m_Q}\ \xi$, and 
$v_h=\sqrt{E_q+m_Q}\ \eta$, we obtain from (38-39), 
\begin{equation}
{\eta}^{\dag}\ 
 \biggl[2E_q\ {\sigma}^j+
\frac{2q^j\ {\bf {\sigma}}{\cdot}{\bf q}}{(E_q+m_Q)} \biggr]\ {\xi}
\end{equation}
i.e. the vertex obtained from full QCD in terms of the Pauli spinors
$\xi$ and $\eta$,  
allowing a systematic non-relativistic expansion: the leading term
$2m(\eta^{\dag}{\bf \sigma}\xi)$ associated with (c) in Eq. (28) as
expected.

Had we started from the Lagrangian ${\cal L}^{(+-)}$, i.e. focusing on a
creation process, the heavy quark vector current would be
\begin{equation}
\overline{u}_h\ \biggl[\frac{2E_q}{E_q+m_Q}{\gamma}^{\mu}_{\bot}+
\frac{2{q'}^{\mu}_{\bot}q{\slash}_{\bot}}{(E_q+m_Q)^2} \biggr]\ v_h
\end{equation}
and since
\begin{equation}
P_+\ q{\slash}_{\bot}\ P_- = 
\left(
     \begin{array}{ll}
      0 & -{\bf \sigma}{\cdot}{\bf q} \\
      0 & \ \ \ 0     
     \end{array}
\right)
\end{equation}
we rapidly recover 
\begin{equation}
{\xi}^{\dag}\ 
 \biggl[2E_q\ {\sigma}^j-
\frac{2q^j\ {\bf {\sigma}}{\cdot}{\bf q}}{(E_q+m_Q)} \biggr]\ {\eta}
\end{equation}
which is the analogous expression
to Eq. (A.11b) in the c.o.m. frame as given in \cite{braaten}.

In sum, we have derived in this Section the vector current of 
on-shell heavy quarks
coupling to a hard gluonic field directly from the 
corresponding Lagrangian  ${\cal L}^{(-+)}$,  ${\cal L}^{(+-)}$
pieces, giving consistency to our formalism.

\section{Heavy Quarkonia Hadroproduction}

In 1992-93, CDF and D0 Collaborations surprisingly
found  an excess of charmonia prompt hadroproduction
at the $p\overline{p}$ Tevatron 
collider \cite{fermi} 
w.r.t. to the theoretical expectations based on what was
considered at that time as conventional wisdom, the so-called
color-singlet model (CSM) \cite{schuler0}. This disagreement was
particularly amazing because data spread over large
transverse momenta where the theoretical analysis ought to be
quite clean. Therefore another mechanism, 
known as the color-octet model (COM), was proposed
by Braaten and Fleming in 1995 \cite{fleming}, superseding
(generalizing actually) the CSM.  Soon later, the COM was
viewed as deriving from NRQCD \cite{bodwin}, giving an adequate
framework for the factorization of the inclusive 
production cross section
into short- and long-distance parts, and providing a sound basis to the
theoretical analysis so far done.

However, current problems in the interpretation of 
charmonia hadroproduction \cite{braaten00}, especially
regarding the (expected but unobserved) transverse 
resonance polarization, 
have cast serious doubts on the validity
of NRQCD when applied to these processes. Actually, the 
meaningful application of the effective theory
relies on the presumed convergence of the expansion
governed by the typical velocity of heavy quarks $v'$
and the strong coupling constant $\alpha_s$.
Different contributions are assumed as leading whereas others
are neglected under the assumption of the validity of the
velocity scaling rules derived from NRQCD, mainly based
on dimensional counting and very general physical arguments. 

In this context some authors \cite{leibovich}
have recently hypothesized that the correct
power counting for charmonia should be the 
dimensionless parameter
${\Lambda}_{QCD}/m_c$, along the lines of
HQET 
\footnote{The authors of \cite{leibovich}
introduce the acronym NRQCD$_c$ which we identify
with HQEFT applied to charmonium.}, leading 
to predictions which could
differ from the expectations coming from the usual velocity
scaling rules of NRQCD.
In the following and based on our previous development,
 we want to give a theoretical
support to such a point of view. In fact
we are advocating in this paper that HQEFT could provide 
an appropriate 
and consistent framework for the
analysis of hadroproduction of heavy quarkonia. 
To this end, from Fig. 1
we may identify the heavy quark three-momentum ${\bf q}$
relative to the c.o.m.
 as the residual three-momentum ${\bf k}$ 
introduced in Section 2. 

On the other hand, in charmonium systems 
it seems sensible to assume 
$v'\ {\simeq}\ {\mid}{\bf q}{\mid}/m_Q$ ${\simeq}\
{\Lambda}/m_Q$, where ${\Lambda}$ stands for
a $\lq\lq$typical'' scale characterizing the quarkonium soft
dynamics, of order few times ${\Lambda}_{QCD}$. Therefore 
dynamical (i.e. non-Coulombic) gluons in $c\overline{c}$ 
bound states should be of the type 
(${\Lambda},{\bf \Lambda}$) though the typical
residual four-momenta of bound heavy quarks should be of the type
(${\Lambda}^2/2m_Q,{\bf \Lambda}$) 
(see \cite{schuler,leibovich}), in close analogy with
heavy-light hadrons. 
Hence the heavy quark field expansions
given in Eqs. (12) and (18) about the heavy quark mass, 
and the reinterpretation of $e^{-i2m_Qv{\cdot}x}A^{\mu}$ in Eq. (29)
as a soft gluonic field coupling to a heavy quark vector current
make sense altogether. 

On the other hand, the effective QCD Lagrangian \cite{bodwin}
to deal with heavy quarkonium is usually written as:
\begin{equation}
{\cal L}_{eff}\ =\ {\cal L}_2\ +\ {\cal L}_4\ +\ {\cal L}_{light}
\end{equation}
where ${\cal L}_2$ contains bilinear operators, ${\cal L}_4$ 
stands for the four-fermion piece and ${\cal L}_{light}$
takes into account light quarks and 
gluons. 

In NRQCD heavy quarks are treated according to a
Schr\"{o}dinger-type field theory with separate two-component
Pauli spinors (commonly denoted as $\psi$ and $\chi$), rather 
than with four-component Dirac fields as
in HQEFT. Relativistic effects of full QCD are reproduced
through correcting terms bilinear in the quark field or
in the antiquark field. However, mixed two-fermion operators
involving both quark and antiquark fields in ${\cal L}_2$
corresponding to creation/annihilation of $Q\overline{Q}$ pairs  
are {\em excluded}.

\begin{figure}[htb]
\centerline{
\epsfig{figure=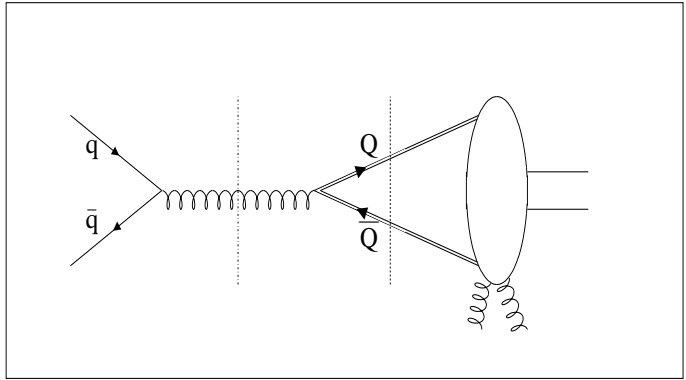,height=5.cm,width=9.cm}}
\caption{\small{Graph representing a $q\overline{q}$ annihilation
followed by the creation of
a $Q\overline{Q}[^3S_1]$ pair, evolving into a final spin-triplet
quarkonium color-singlet state 
by emission of two soft gluons via a double
chromoelectric dipolar transition. The two dotted lines 
delimit the gluon-quark vertex 
to be described along the lines of HQEFT
following our suggestion of including the description
of the $Q\overline{Q}$ creation within the realm
of HQEFT. Moreover, local four-fermion terms in the Lagrangian
accounting for heavy quarkonium production can
be thought as derived from the ${\cal L}^{(+-)}$
and ${\cal L}^{(-+)}$ pieces of
the same HQEFT Lagrangian.}}
\end{figure}

On the other hand, in order to account for
heavy quarkonium annihilation or creation, a set of four-fermion
operators are thereby included in the piece ${\cal L}_4$ of the
effective Lagrangian (44), representing local
interactions whose short-distance coefficients have
to be determined by means of a matching
procedure to QCD as described, for instance, in \cite{braaten}.

Actually, in this work we are advocating that
the ${\cal L}_2$ piece of the effective Lagrangian
can be extended as
\begin{equation}
{\cal L}_2\ =\ {\cal L}_2^{(++)}+{\cal L}_2^{(--)}+{\cal L}_2^{(-+)}+
{\cal L}_2^{(+-)}
\end{equation}
where the ${\cal L}_2^{(-+)}$ and ${\cal L}_2^{(-+)}$ terms
can be identified with those coming from Eqs. (26-27), 
therefore taking into account $Q\overline{Q}$ pair formation
and annihilation at tree-level. Notice that, as far as
the heavy quark mass dependence has been
removed from the effective fields, only low-energy modes
(of the order of ${\Lambda}$ around $m_Q$,) remain as stressed before.
Hence the situation turns out to be very similar to
the standard application of HQEFT to heavy-light systems.
On the other hand, we keep the 
${\cal L}_4$ piece in the effective Lagrangian, although
written in terms of HQEFT fields.

For definiteness, let us consider quarkonium hadroproduction
through the partonic channel: 
\[ q\overline{q}\ {\rightarrow}\ g^{\ast}\ {\rightarrow}\ 
(Q\overline{Q})_8[^3S_1]\ {\rightarrow}\ H\ X \] 
where $H$ denotes a particular quarkonium state. (The
diagrammatic representation for this production process yielding
a vector resonance is shown in Fig. 2.)

\subsection{Color-octet matrix elements from HQEFT}

Under the assumption of factorization, the
cross section can be written  
for inclusive resonance production as \cite{bodwin}
\begin{equation}
d\sigma(H+X)\ =\ \sum_{n}\ d\hat{\sigma}[(Q\overline{Q})_n+X]\ <{\cal{O}}_n^H>
\end{equation} 
where $d\hat{\sigma}$ stands for the production cross section
of a $Q\overline{Q}$ pair in a definite color and angular-momentum 
state labeled by $n$. The short-distance interaction, represented 
by $d\hat{\sigma}$ in (46), can be accounted for by 
perturbative QCD, whereas the long-distance process is encoded
in the vacuum expectation values of four-fermion local operators, 
i.e. $<{\cal{O}}_n^H>$ \cite{bodwin}. According to NRQCD, the relative
importance of the various terms in the factorization 
is determined by the order in $\alpha_s$ and some dimensionless
ratios of kinematic variables 
in the short-distance factors, and by a hierarchy 
based on the order in $v'$ of
the matrix elements. Hence, the velocity scaling of the MEs
is determined by the number of derivatives in the respective
operators and the number of electric and magnetic
dipole transitions needed to reach the final physical
quarkonium state from the initial $Q\overline{Q}$ pair
produced at the short-distance process.

Pictorially, the long-distance evolution lies on the right 
of the rightmost dotted line in Fig. 2, belonging to the realm of the effective 
Lagrangian ${\cal L}_4$ and providing
the probability for a heavy quark-antiquark pair to 
evolve into a (definite) final quarkonium state. 
On the other hand, the 
subprocess $g^{\ast}\ {\rightarrow}\ Q\overline{Q}$ can be 
ascribed to the piece  ${\cal L}_2^{(+-)}$ of the effective Lagrangian
from our viewpoint.
Therefore we shall use the 
${h}_v^{(+)}$ and ${h}_v^{(-)}$ HQEFT fields when rewriting the
relevant operators. For example, 
the $O_8(^3S_1)$ 6-dimensional operator \cite{bodwin} can be 
expressed in
the form of a current${\times}$current term
(see figure 3), i.e.
\begin{equation}
O_8(^3S_1)\ =\ \bar{h}_v^{(-)}\ {\gamma}_{\bot}^{\mu}\ T_a\ h_v^{(+)}\ 
(a_H^{\dag}a_H)\ \bar{h}_v^{(+)}\ {\gamma}_{\bot\mu}\ T_a\ h_v^{(-)}
\end{equation}
where $a_H$($a_H^{\dag}$) stands for annihilation (creation) operators
of heavy quarkonium $H$, such that
\begin{equation}
a_H^{\dag}a_H\ =\ \sum_{X}{\mid}H+X><H+X{\mid}
\end{equation}

The above $O_8(^3S_1)$ operator can be
seen as deriving from the corresponding piece
of the ${\cal L}_4$ Lagrangian \cite{bodwin}
now expressed in terms of HQEFT fields,
\begin{equation}
{\cal L}_4^{^3S_1^{[8]}}\ =\ \frac{F_8(^3S_1)}{m_Q^2}\ \
\bar{h}_v^{(+)}{\gamma}_{\bot}^{\mu}\ T_a\ h_v^{(-)}\ 
 \bar{h}_v^{(-)}{\gamma}_{\bot\mu}\ T_a\ h_v^{(+)}
\end{equation}

The physical interpretation of (47) is that of a local four-fermion
operator which creates a slightly off-shell $Q\overline{Q}$ pair
in a $^3S_1^{[8]}$ state 
and destroys it again, analogously as in NRQCD.
Its vacuum expectation value
 $<0{\mid}O_8(^3S_1){\mid}0>{\equiv}<O_8(^3S_1)>$ corresponds to 
the usual color-octet matrix element of NRQCD.  
Notice, however, that the current${\times}$current structure can be thought 
now as
deriving from the ${\cal L}_2^{(-+)}$ and ${\cal L}_2^{(+-)}$ Lagrangian 
pieces, involving only low-frequency modes and providing consistency to
all the theoretical framework. 
High-energy modes are encoded in the short-distance 
coefficients ($F_8(^3S_1)$ in this particular case), as usual in 
low-energy effective theories. (Nevertheless, the matching procedure,  
although formally the same as in NRQCD,    
might be seen as  $\lq\lq$built in'' because the effective fields
in the four-quark local operators follow 
from ${\cal L}_2$.) 

Hence the power counting of the
effective theory should follow the lines of HQEFT, as far as
$m_Qv'\ {\simeq}\ {\Lambda}$, and the expansion
parameter can be identified as ${\Lambda}/m_Q$.
However, let us remark that
the intermediate colored bound state $(Q\overline{Q})_8$ evolving into
final charmonium actually is a kind
of {\em hybrid} state \cite{grinstein98}, whose $\lq\lq$size'' 
(and even nature) is
not accurately known. Therefore, one could expect ${\Lambda}$ to be
of order few times ${\Lambda}_{QCD}$, namely 700-800 MeV, and 
typically $({\Lambda}/m_Q)^2\ {\simeq}\ 1/4$, but
with large uncertainties.
\vskip 1.cm

\begin{figure}[htb]
\centerline{\hbox{
\epsfig{figure=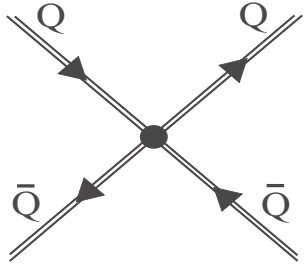,height=3.5cm,width=4.cm}}}
\caption{\small{ Since the formation
of the quarkonium state $H$ takes place over distances of
the order of $1/{\Lambda}$, the use of HQEFT is justified. 
Local interactions give rise to the $O_8(^3S_1)$ operator
among others, 
in terms of ${h}_v^{(\pm)}$ fields coming from the 
${\cal L}_2$ piece of the Lagrangian.}}
\end{figure}

\subsection{Power scaling rules}

Under the assumption that HQEFT could be a suitable effective
theory to deal with charmonia production, we will follow a
reasoning analogous to \cite{leibovich,schuler}, arguing that there 
could be a hierarchy of the matrix elements, different 
from $\lq\lq$standard'' NRQCD. Therefore we expect that
a single chromoelectric dipole (E1) transition  should
scale as ${\Lambda}/m_Q$ {\em at the amplitude level}.
Since two E1 transitions are required to reach a
$^3S_1^{[1]}$ color-singlet state
from a $^3S_1^{[8]}$ color-octet one, the 
resulting suppression factor
is $({\Lambda}/m_Q)^2$. On the other hand,  
a single chromomagnetic transition (M1) should 
scale as ${\Lambda}/m_Q$, because of the time derivative
on the gluonic field strength-tensor, 
picking up a typical energy of order ${\Lambda}$. Thus,  
{\em at the cross section level}, the 
power countings are expected to be
of order $({\Lambda}/m_Q)^4$ and $({\Lambda}/m_Q)^2$
for a double E1 transition and a single M1
transition, respectively.

Now, defining as usual the
linear combination of MEs 
\footnote{The little difference in shape between the $^1S_0^{[8]}$ and 
$^3P_J^{[8]}$ contributions as a function of transverse momentum
does not permit independent fits and
only a linear combination of them can be extracted from
experimental data \cite{cho,mas97a}.}
(where $r$ is an adjustable  parameter
varying between 3 and 3.5 for charmonia hadroproduction) as
\begin{equation}
M_r^{\psi(nS)}\ =\ <O_8^{\psi(nS)}(^1S_0)>\ 
+\ r\ \frac{<O_8^{\psi(nS)}(^3P_J)>}{m_Q^2}
\end{equation}
the former one, i.e. $<O_8^{\psi(nS)}(^1S_0)>$, should 
become leading in the overall power counting of $M_r^{\psi(nS)}$.

Let us now define the ratio
\begin{equation}
R^{\psi(nS)}\ =\ \frac{M_r^{\psi(nS)}}{<O_8^{\psi(nS)}(^3S_1)>}\ \ \ \ 
\ ;\ \ \ (n=1,2)
\end{equation}

Since HQEFT predicts a typical scaling (relative to the
color-singlet contribution) of
$({\Lambda}/m_Q)^4$ for $<O_8^{\psi(nS)}(^3S_1)>$, 
and $({\Lambda}/m_Q)^2$ for $M_r^{\psi(nS)}$, one should expect
\begin{equation}
R^{\psi(nS)}\ {\sim}\ \biggl(\frac{m_Q}{\Lambda}\biggr)^2
\end{equation}
which could also be expressed as $1/{v'}^2$ in terms of the
more conventional power counting based on the typical velocity
of heavy quarks, under the assumption that 
${v'}\ {\simeq}\ {\Lambda}/m_Q$. 

Conversely, in standard NRQCD
both matrix elements $<O_8^{\psi(nS)}(^3S_1)>$ and $M_r^{\psi(nS)}$,   
should scale similarly  (as ${v'}^4$ w.r.t. to the
leading color-singlet component). Therefore  no
hierarchy should appear, and 
one would expect $R^{\psi(nS)}\ {\simeq}\ 1$.

In the following we carry out a check based on the available
experimental information 
\footnote{We limit ourselves to those references
where MEs extractions of {\em both} $J/\psi$ and $\psi'$ are
given; notice that distinct
theoretical inputs may have been used, as for example different
parton distribution functions.} extracted by several authors from 
fits to Tevatron data on charmonia hadroproduction.
In tables 1 and 2 we show the values of the
color-octet matrix elements $<O_8^{\psi(nS)}(^3S_1)>$
and $M_r^{\psi(nS)}$ for $J/\psi$ and $\psi'$ respectively, 
and the corresponding ratios $R^{\psi(nS)}$.

\begin{table*}[hbt]
\setlength{\tabcolsep}{1.5pc}
\caption{Values (in units of $10^{-3}$ GeV$^3$)
of $<O_8^{\psi(1S)}(^3S_1)>$
and $M_r^{\psi(1S)}$ matrix elements ($r$ varies between 3 and 3.5) 
and their ratios $R^{\psi(1S)}$, obtained 
from Tevatron data on prompt 
$J/\psi$ inclusive hadroproduction. Error bars are only statistical.}
\label{FACTORES}

\begin{center}
\begin{tabular}{ccccc}    \hline
ME: & $<O_8^{\psi(1S)}(^3S_1)>$  & $M_r^{\psi(1S)}$ & $R^{\psi(1S)}$ \\
\hline
\cite{nason00} & $11.9{\pm}$1.4 & $45.4{\pm}11.1$ & $3.8{\pm}1.0$  \\
\hline
\cite{beneke97} & $10.6{\pm}1.4$ & $43.8{\pm}11.5$ & $4.1{\pm}1.2$  \\
\hline
\cite{mas97a} & $3.3{\pm}0.5$ & $14.4{\pm}2.8$ & $4.4{\pm}1.1$   \\
\hline
\cite{cho} & $6.6{\pm}2.1$ & $66{\pm}5$ & $10.0{\pm}3.3$  \\
\hline
\cite{braaten00} & $3.9{\pm}0.7$ & $66{\pm}7$ & $16.9{\pm}3.5$  \\
\hline
\end{tabular}
\end{center}
\end{table*}

\begin{table*}[hbt]
\setlength{\tabcolsep}{1.5pc}
\caption{Values (in units of $10^{-3}$ GeV$^3$)
of $<O_8^{\psi(2S)}(^3S_1)>$
and $M_r^{\psi(2S)}$ matrix elements ($r$ varies between 3 and 3.5) 
and their ratios $R^{\psi(2S)}$, obtained 
from Tevatron data on prompt 
${\psi}'$ inclusive hadroproduction. Error bars are only statistical.}
\label{FACTORES}

\begin{center}
\begin{tabular}{cccc}    \hline
ME: & $<O_8^{\psi(2S)}(^3S_1)>$ & $M_r^{\psi(2S)}$ & $R^{\psi(2S)}$ \\
\hline
\cite{braaten00} & $3.7{\pm}0.9$ & $7.8{\pm}3.6$ & $2.1{\pm}1.1$  \\
\hline
\cite{mas97a} & $1.4{\pm}0.3$ & $3.3{\pm}0.9$ & $2.4{\pm}0.8$   \\
\hline
\cite{nason00} & $5.0{\pm}0.6$ & $18.9{\pm}4.6$ & $3.8{\pm}1.0$  \\
\hline
\cite{cho} & $4.6{\pm}1.0$ & $17.7{\pm}5.7$ & $3.8{\pm}1.5$  \\
\hline
\cite{beneke97} & $4.4{\pm}0.8$ & $18.0{\pm}5.6$ & $4.1{\pm}1.5$  \\
\hline
\end{tabular}
\end{center}
\end{table*}

We realize that indeed a hierarchy seems to exist
for  $<O_8^{\psi(nS)}(^3S_1)>$
and $M_r^{\psi(nS)}$. Although $R^{\psi(1S)}$ 
varies in a somewhat wide range, a value equal or greater than 
{\em four}
seems to be favored.
On the other hand, the ratio 
$R^{\psi(2S)}$ turns out to be slightly but 
systematically smaller than  
$R^{\psi(1S)}$, but close to {\em four} as well, i.e.
compatible with the expectation
${v'}^2\ {\simeq}\ {\Lambda}^2/m_Q^2\ {\simeq}\ 1/4$, already
mentioned at the end of Section 6.1. 

In fact one should  bear in mind
that t$R^{\psi(nS)}$ is not a fixed quantity for the
whole charmonium family (nor the velocity $v'$).  
This remark is in agreement with the expected larger value
of the typical velocity $v'$ for higher $n$ states in
the charmonium sector or, equivalently, a smaller
ratio $m_Q/{\Lambda}$.

Another source for the determination of long-distance 
matrix elements comes
from fixed-target experiments. In Ref. \cite{beneke96}
Beneke and Rothstein find $M_7^{\psi(1S)}=30$, having fixed 
$<O_8^{\psi(2S)}(^3S_1)>$
equal to 6.6 (in units of GeV$^{-3}$). For the $\psi'$ resonance, the
numerical values are $M_7^{\psi(2S)}=5.2$ having fixed 
$<O_8^{\psi(1S)}(^3S_1)>$
equal to 4.6 (also in units of GeV$^{-3}$)
\footnote{In fixed-target experiments it is not
possible to fit simultaneously both matrix
elements because they display a similar shape as
a function of the beam energy.}. Therefore one gets
$R^{\psi(1S)}=4.5$ and $R^{\psi(2S)}=1.1$, indicating again a 
smaller value for the $2S$ state, although subject to 
large uncertainties too
\footnote{In Ref. \cite{maltoni99} smaller values for
$M_7^{\psi(nS)}$ are obtained as the inclusion of NLO corrections
tend to decrease the values needed for the MEs. However, 
all values shown in tables 1 and 2 refer to LO extractions
so we prefer not mixing them.}.

NRQCD matrix elements can also be determined
from inclusive decays of $B$ mesons. In a recent
work \cite{ma00}, Ma obtains a combination of $<O_8^{\psi(nS)}(^1S_0)>$ and
$<O_8^{\psi(nS)}(^3P_J)>$ MEs in the framework of HQEFT, having
fixed $<O_8^{\psi(nS)}(^3S_1)>$ equal to $10.6(4.4){\cdot}10^{-3}$ GeV$^3$
for the $J/\psi$ and $\psi'$ respectively.
The resulting ratios 
(not exactly the same as in tables 1 and 2 since
$r$ takes the value of 1.13) are $R^{\psi(1S)}\ {\simeq}\ 
R^{\psi(2S)}\ {\simeq}\ 2.3$ which, although slightly smaller than
those presented in the tables, are not inconsistent
with them in view of the smaller $r$ value and the uncertainties 
involved in the extraction \cite{ma00}.

\section{Summary and last remarks}

In this paper we have carefully reviewed all steps to
derive from the full QCD tree-level Lagrangian 
a complete transformed Lagrangian in 
terms of the heavy quark effective fields $h_v^{(\pm)}$, keeping the 
particle-antiparticle mixed pieces allowing for heavy quark-antiquark 
pair annihilation/creation. Let us note that such pieces are not
generally neither used nor shown in similar developments in the 
literature, with a few exceptions \cite{wu93,mas99,wu,wu1}.

Indeed, it may seem quite striking that a low-energy effective theory 
could be appropriate to deal
with hard processes such as $Q\overline{Q}$ annihilation or creation. 
The keypoint is that, assuming a kinematic regime where heavy quarks/antiquarks
are almost on-shell and moving with small relative momentum, 
the strong momentum dependence associated with the heavy
quark masses can be removed, so that a description 
based on the low frequency
modes of the fields still makes sense, 
as in HQET applied to heavy-light hadrons. Such a kinematic regime can
be well 
matched by heavy quarkonia and intermediate colored
$(Q\overline{Q})_8$ bound states predicted by the COM.

In particular, we have focused on an annihilation process with initial-state 
quarks satisfying the Dirac equation of motion for free fermions. Thus, 
we have derived directly from the ${\cal L}^{(-+)}$ piece the heavy quark
vector current coupling to a background gluonic field, recovering a well known
expression shown in the literature \cite{braaten} allowing
a non-relativistic expansion in the matching procedure of NRQCD and
full QCD. 

In the second part of this work we have examined the meaning
of using a HQEFT Lagrangian to account for heavy quarkonium
production. Thereby we kept the ${\cal L}^{(-+)}$ and 
${\cal L}^{(+-)}$ pieces in the effective Lagrangian, from 
which the ${\cal L}_4$ four-fermion
piece can be $\lq\lq$constructed''. The result 
looks formally the same as in conventional NRQCD
although the fields are the HQEFT ones, i.e. $h_v^{(\pm)}$, giving
self-consistency to all the framework.

From a phenomenological point of view, we advocate 
the possibility of using the
extended version of HQEFT for describing inclusive hadroproduction of 
charmonium states according to the color-octet mechanism. Indeed, the
four different energy scales usually invoked in NRQCD (i.e. $m_Q$,
$m_Qv'$, $m_Q{v'}^2$ and ${\Lambda}_{QCD}$) basically reduce to two
in HQEFT: $m_Q$ and a typical hadronic scale, $\Lambda$, of the order
of a few ${\Lambda}_{QCD}$'s, in a reasonable accordance
with the dynamics of charmonium systems. 
Conversely, inclusive
production of bottomonium
resonances should likely be better described by NRQCD because of the 
clearer difference among the scales.

From inspection of
tables 1 and 2, we can conclude that there are hints indicating
that HQEFT may be considered as a candidate
for describing $J/\psi$ and $\psi'$ hadroproduction - although likely
in a limited way because of the approximations in the scales and the
lack of a static limit {\em \`a la HQET} when 
${\Lambda}/m_Q\ {\rightarrow}\ 0$, 
as stressed elsewhere \cite{mas95,leibovich}.
Hence we give support to the hypothesis
presented in Ref. \cite{leibovich} of basing the power counting 
for charmonium hadroproduction along
the lines of HQEFT, to some extent.

As a final comment, let us remark that still the uncertainties
in the phenomenological extractions of the color-octet
matrix elements (choice of numerical values
for the heavy quark mass and renormalization/fragmentation scales, parton 
distribution functions, etc) are quite
large. Moreover, new approaches have recently appeared (e.g. the
$k_T$ factorization \cite{hagler,chao}) while 
the experimental evidence is not yet conclusive to decide 
which framework should be the most
adequate to describe hadroproduction
of heavy resonances. 
\newline

\subsection*{Acknowledgements} I thank 
comments and discussions with J.L. Domenech, O. Teryaev and S. Wolf.

\end{document}